\documentclass[usenatbib,fleqn,12pt]{wlscirep}

\usepackage{graphicx}
\usepackage{amssymb}
\usepackage{newfloat}
\DeclareFloatingEnvironment[name={Extended Data Figure}]{extfig}
\usepackage{soul}
\newcommand{\apj}{Astrophys. J.}           % Astrophysical Journal
\newcommand{\apjl}{Astrophys. J.}           % Astrophysical Journal
\newcommand{\mnras}{Mon. Not. R. Astron. Soc.}       % Monthly Notices of the RAS
\newcommand{\nat}{Nature}
\newcommand{\aap}{Astron. Astrophys.}
\newcommand{\araa}{Annual Rev. Astron. Astrophys.}
\newcommand{\aj}{Astron. J.}
\newcommand{\pasp}{Pubbl. Astron. Soc. Pacific}
\title{The birth of the Milky Way as uncovered by accurate stellar ages with Gaia}

\author[1,2,*]{Carme Gallart}
\author[3]{Edouard J. Bernard}
\author[1,2]{Chris B. Brook}
\author[1,2]{Tom\'as Ruiz-Lara}
\author[4,5]{Santi Cassisi}
\author[3]{Vanessa Hill}
\author[1,2]{Matteo Monelli}

\affil[1]{Instituto de Astrof\'isica de Canarias, E-38200 La Laguna, Tenerife, Spain}
\affil[2]{Departamento de Astrof\'isica, Universidad de La Laguna, E-38205 La Laguna, Tenerife, Spain}
\affil[3]{Universit\'e C\^ote d'Azur, Observatoire de la C\^ote d'Azur, CNRS, Laboratoire Lagrange, France}
\affil[4]{INAF -- Astronomical Observatory of Abruzzo, via M. Maggini, sn, 64100 Teramo, Italy}
\affil[5]{INFN, Sezione di Pisa, Largo Pontecorvo 3, 56127 Pisa, Italy}
\affil[6]{Laboratoire Lagrange, Universit\'e de Nice Sophia-Antipolis, Observatoire de la C\^ote d'Azur, CNRS, Bd de l'Observatoire, CS 34229, F-06304 Nice cedex 4, France}
\affil[*]{carme@iac.es}

\begin{document}

\flushbottom
\maketitle

\thispagestyle{empty}
\vspace{-1cm}

{\bf Knowledge of ages for stars formed over a galaxy's lifetime is fundamental to understand its formation and evolution. However, stellar ages are difficult to obtain since they cannot be measured from observations, being comparison with stellar models\cite{soderblom2010} required.  Alternatively, age distributions can be derived applying the robust technique of colour-magnitude diagram fitting\cite{gallart2005}, till now mainly employed to study nearby galaxies. The new distances to individual Milky Way stars from the Gaia mission\cite{brown2018} have allowed us to use this technique to derive ages from a thick disk colour-magnitude diagram, and from the enigmatic, two-sequenced colour-magnitude diagram of the kinematically hot local halo\cite{babusiaux2018}, which blue-sequence has been linked to a major accretion event\cite{haywood2018, helmi2018}. Because accurate ages were lacking, the time of the merger and its role on our Galaxy's early evolution remained unclear. We show that the stars in both halo sequences share identical age distributions, and are older than the bulk of thick disc stars. The sharp halo age cut 10 Gyr ago can be identified with the accretion of Gaia-Enceladus.  Along with state-of-the-art cosmological simulations of galaxy formation\cite{brook2012magicc}, these robust ages allow us to order the early sequence of events that shaped our Galaxy,  identifying the red-sequence as the first stars formed within the Milky Way progenitor which, because of their kinematics, can be described as its long sought in-situ halo. }

%which is naturally more metal rich than the accreted halo population originating in the lower-mass galaxy Gaia-Enceladus.}
%These findings provide a crystal clear picture of the early sequence of events that gave rise to the complex Milky Way structure, and have far reaching implications for understanding the formation of disc galaxies in general.}

\vspace{1.0 cm}

%As Milky Way-like galaxies form, smaller stellar systems accrete onto the main galactic progenitor, assembling a kinematically hot, spheroidal-like stellar halo surrounding the emerging galactic disc\cite{searle_zinn1978, bullock2005}. Recently,  it has become clear that the Milky Way experienced a very significant merging event early in its formation\cite{helmi2018}, of which the first hints date back to at least 15 years ago\cite{brook2003, nissen_schuster2010, schuster2012, belokurov2018}. Here, we will derive  precise age distributions of the stars in the two enigmatic sequences discovered in the CMD of the kinematically hot halo stars which distances and luminosities were provided in the second data release of the Gaia mission (DR2)\cite{babusiaux2018}. These ages, and their comparison with those inferred from the CMD of the thick disk, provide crucial constraints on the timing and the nature of this galaxy merger, and on its role in the early evolution of the Galaxy\cite{helmi2018}.

The new accurate parallaxes and luminosities provided in the second data release (DR2) of the Gaia mission\cite{brown2018} have allowed us to construct, for the first time, colour-magnitude diagrams (CMD) in the {\it absolute} plane for stars located in a large volume of the Milky Way, encompasing different Galactic structural components. These CMDs, in units of absolute magnitudes and colours, are what is required to derive star formation histories\cite{gallart2005} and stellar age distributions, by comparing them with theoretical CMDs derived from stellar evolution models\cite{pietrinferni2006}. 

\begin{figure}[h]
\begin{center}
\includegraphics[width=0.8\textwidth]{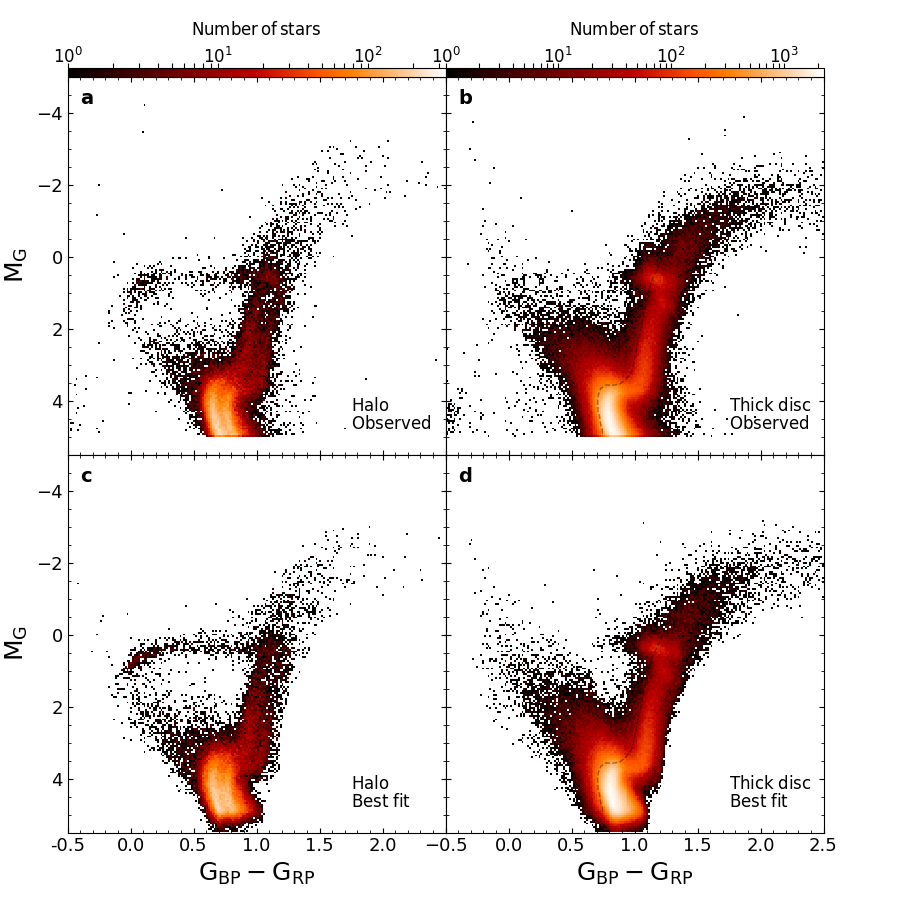} 
\end{center}
\caption{{\bf Milky Way halo and thick disc observed and modelled colour-Magnitude Diagrams (CMDs).} Panels {\bf a} and {\bf b} display the observed Gaia CMDs for the halo and thick disc sub-populations, respectively. They have been represented as Hess diagrams with a logarithmic normalization of the number of stars in the colour-magnitude bins. Two sequences are clear in panel {\bf a}: the blue sequence on the left and the red sequence on the right. Panels {\bf c} and {\bf d} display the best-fit models to the halo and thick disc CMDs, respectively.  The solutions for the fit using $\alpha$-enhanced models have been represented in this Figure, see Methods. The solutions obtained using the solar scaled models are not noticeably different. The dashed line in panels {\bf b} and {\bf d} indicates the position of the gap between the blue and red halo sequences. See Methods for more information on the sample selection and the CMD fitting technique.}
\label{fig:plot1}
\end{figure}

The top panels of Figure~\ref{fig:plot1} show the CMD of two sub-populations of Milky Way stars taken from a parent population that lies within a sphere of $\simeq$ 2 Kpc around the Sun, as observed by Gaia. In this volume, accurate distances and absolute magnitudes can be derived directly from parallaxes. The  CMD in the top left panel contains about sixty thousand stars from this spherical region with large tangential velocities relative to the Sun (greater than 200 km/s). Stars with such high velocities are  classified in this study as belonging to a kinematically defined stellar halo\cite{babusiaux2018}. The CMD in the top right panel is of some half million stars from the same spherical region but selected  to be at least 1.1 Kpc above or below the Galactic plane. At this distance from the plane, the majority of stars are expected to belong to the thick disc\cite{gilmore_reid1983}, rather than to the young thin disc component. We have excluded from this sample the stars with high velocity (greater than 200 km/s) that have been included in the halo CMD.  Note that our thick disk definition is morphological rather than chemical or kinematical: some stars with thin disk kinematics \cite{babusiaux2018} would be included in our sample. For simplicity, in this paper we will refer to the two samples we have just defined as {\it halo} and {\it thick disk} (see Methods for further information on the sample selection criteria).  As  is readily apparent, the CMD of the kinematically hot halo population contains two distinct sequences: a ''blue'' sequence on the left, and a ''red'' sequence on the right. This peculiar morphology indicates the presence of two distinct sub-populations within the halo.  The blue sequence has been associated to a very significant merging event, Gaia-Enceladus, occured early during the Milky Way formation\cite{helmi2018}, and of which the first hints date back to at least 15 years ago\cite{brook2003, nissen_schuster2010, schuster2012, belokurov2018}. The nature of the red sequence has been less clear. It has been associated with the Milky Way thick disc \cite{haywood2018}, with support for this association coming from the characteristics of its chemical composition \cite{schuster2012}.

We have compared the distribution of stars in these observed CMDs with that in CMDs computed with state-of-the-art stellar evolution models \cite{pietrinferni2006} and found the combination of ages and chemical composition of the stars (or metallicities, that is, amount of chemical elements other than Hydrogen and Helium) that best reproduce the observed CMDs (see Methods). These best-fit CMDs are displayed in the bottom two panels of Figure~\ref{fig:plot1}.  The model CMDs clearly succeed at reproducing the various features observed in the empirical CMDs.

\begin{figure}[h]
\begin{center}
\includegraphics[width=0.8\textwidth]{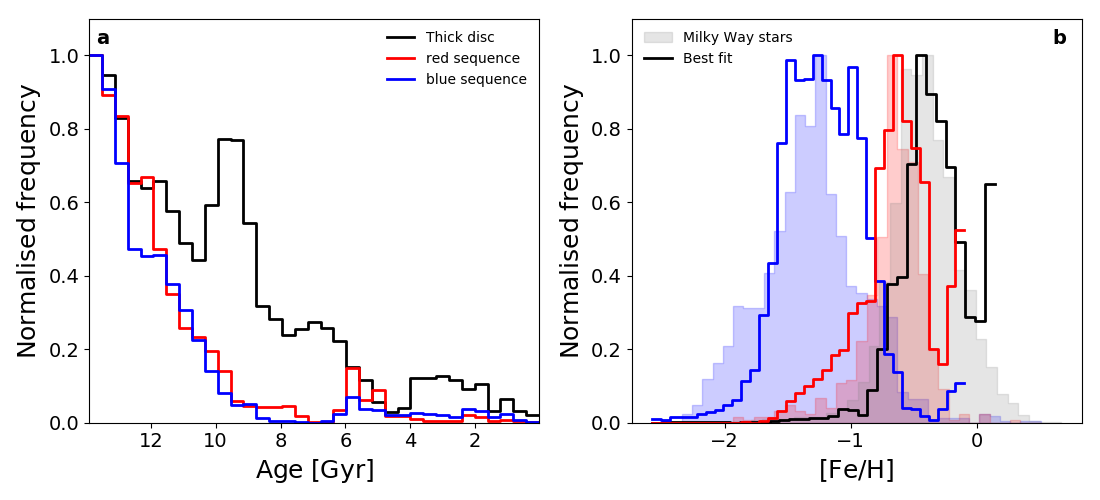} 
\end{center}
\caption{{\bf Milky Way halo and thick disc stellar age and iron [Fe/H] distributions.} Panel {\bf a} shows the distribution of stellar ages of the thick disc (grey histogram), blue (blue histogram) and red (red histogram) halo sequences according to their best-fit model CMDs adopting the $\alpha$-enhanced stellar evolution models. The solutions obtained using the solar scaled models show similar features. Panel {\bf b} shows the distribution of [Fe/H] of the best-fitting models for the blue and red halo sequences as well as the thick disc stars  (solid lines), compared to [Fe/H] distributions for the stars in each sample that had measured [Fe/H] in the LAMOST or GALAH spectroscopic surveys\cite{zhao2012, buder2018, sanders2018} (filled histograms). The fair agreement between the [Fe/H] distributions inferred with the CMD fitting technique and those measured spectroscopically provides an external check of the reliability of the results.}
\label{fig:plot2}
\end{figure}

Figure~\ref{fig:plot2} displays the distribution of ages and metallicities of these best-fit model CMDs, divided in the blue and red sequences in the case of the halo CMD. The age distributions clearly demonstrate that {\it the two sequences in the halo CMD are composed by stars that are coeval and formed at the earliest possible times in the life of the Universe} (peak age 13.4 Gyr, 50\% of stars formed by 12.3 Gyr ago). The difference in colour is a result of the red and blue sequence stars having different metallicities. The models also clearly show that the thick disc population spans a wider age range than the two halo populations: its stars closely follow the age distribution of the halo for the first $\simeq$ 2 Gyr, but the thick disk keeps forming stars after this initial period and until $\simeq$ 6 Gyr ago, with its age distribution reaching a second peak later on ($\simeq$ 9.5 Gyr ago, with 50\% of stars formed by 10.5 Gyr ago). 

These age results contrast with literature values for the ages  \cite{schuster2012, hawkins2014, Ge2016} assigned to stars associated with these components on which previous conclusions on their nature relied\cite{helmi2018}. These former age determinations indicated that the two halo populations had different ages, with blue-sequence stars about 1-2 Gyr younger than stars in the red-sequence. However, they were based either on i) a relatively small number of stars, and used spectroscopically determined stellar parameters which allow less robust age estimates than absolute luminosities and colours or ii) indirect age estimates based on low resolution spectroscopy. They constrasted with ages inferred with simple isochrone fitting taking into account average metallicities for stars in the two sequences\cite{helmi2018,babusiaux2018}. Our robust age distributions for the thick disk and halo components provide key insights into the first events that shaped the Milky Way structure, providing crucial new evidence that complements previous studies  which have largely dealt with chemical and kinematic information \cite{hawkins2015, bonaca2017, hayes2018, fernandez_alvar2018, haywood2018, helmi2018, mackereth2019, DiMatteo2018arXiv181208232D}.
 
\begin{figure}[h]
\begin{center}
\includegraphics[width=0.8\textwidth]{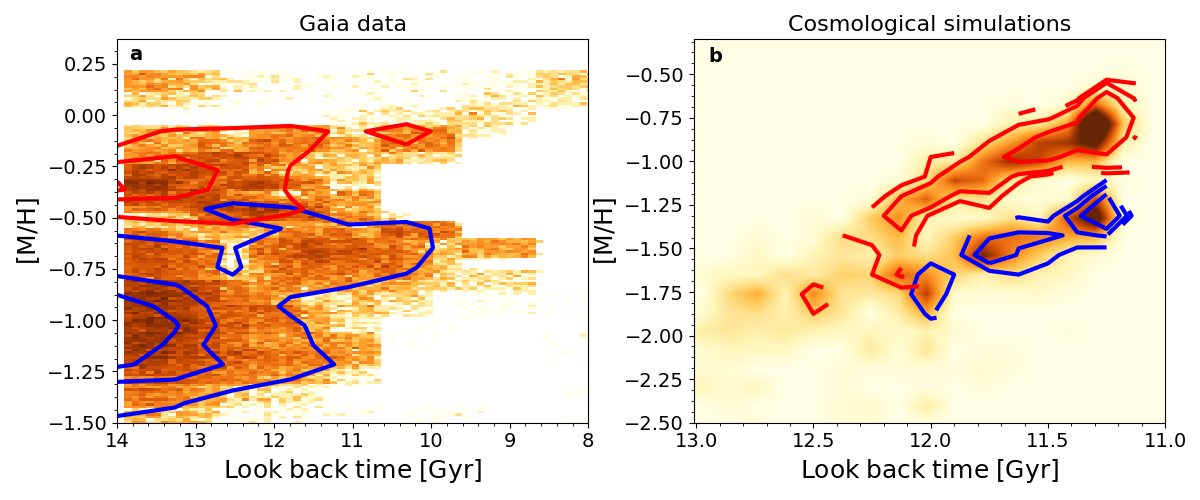} 
\end{center}
\caption{{\bf Age-metallicity relations for the two halo populations as compared to a simulated Milky Way analogue.} Panel {\bf a}: age-metallicity relation for the best-fit model of the halo CMD. The blue and red contours outline the distribution of stars in the blue and red halo sequences, respectively, overlaid to the general distribution of halo stars. Panel {\bf b}, same as {\bf a} but for a simulated Milky Way analogue (see Methods). Blue and red contours depict the age-metallicity distribution of simulated stellar particles belonging to an accreted galaxy (analogue to Gaia-Enceladus) and the Galaxy progenitor, respectively.  The last major merger can be identified as two tracks in the age-metallicity plot, with the higher metallicity stars belonging to the main progenitor and the lower metallicity stars coming from the merged satellite. The agreement, only expected to be qualitative, is remarkable. The time and mass of the last major merger, which can be seen to differ between the simulation and the real Milky Way, determine the details of the ages and metallicities of the red and blue halo populations. 
}
\label{fig:plot3}
\end{figure}

The fact that the two halo populations are coeval means not only that the red and blue populations formed stars at similar times, but that they also stopped forming at similar times. Additionally,  the well established relation \cite{erb2006} between a galaxy's mass and the amount of metals (defined as chemical elements other than Hydrogen and Helium) contained in their stars means that the stars in the red sequence of the halo, being more metal rich, must have formed in a galaxy that was more massive than the one where the stars in the blue sequence were formed. Both these findings are precisely what is expected if these two populations were involved in a merger event, with the red sequence stars belonging to the main progenitor of the Milky Way, and the blue sequence belonging to a smaller accreted galaxy, the one dubbed Gaia-Enceladus. This is shown in Figure~\ref{fig:plot3}. The left panel shows the age-metallicity distribution derived from the observed CMDs, with the red and blue sequence halo stars indicated as red and blue contours. The right panel shows the signature in the age-metallicity plane left  by a merger event in a simulated Milky-Way analogue galaxy (see Methods), where the main progenitor stars are shown as red contours and those from the accreted galaxy as blue contours. The merger can be identified as two tracks in this plot, with the higher metallicity stars belonging to  the main progenitor and the lower metallicity stars coming from the less massive merged galaxy.  

The difference in the amount of metals in the red and blue sequences indicates that the accreted Gaia-Enceladus  had about 30\% \cite{ma2016} of the mass of stars in the main progenitor, although we stress that this ratio remains quite uncertain.  This would indicate a total mass ratio of around 4:1 between the two galaxies, given the relation between stellar mass and total mass \cite{behroozi2013}. This estimated mass ratio is similar to that derived in previous works \cite{helmi2018} using different arguments. Regardless of the exact mass ratio, this encounter heated some of the main progenitor stars that had been forming in a disc-like structure, to the extreme kinematics that lead them to be classified as halo stars. The existence of such halo stars has been predicted by cosmological simulations of Milky Way-type galaxies \cite{zolotov2009, bonaca2017} and has also been postulated from observational evidence \cite{nissen_schuster2010, schuster2012, haywood2018}. {\it Our age determination discloses that the stars heated to halo-like kinematics were among the first formed in the Milky Way}, during the first $\simeq$ 3 Gyr of its evolution, right before the merger with Gaia-Enceladus took place. We can date the merger as occurring about 10 Gyr ago.

The similar star formation timescales for the two halo sequences also has important implications for the origin of their detailed chemical enrichment. It is known \cite{schuster2012, haywood2018, helmi2018, babusiaux2018} that for stars with similar amounts of Fe, the blue sequence stars (Gaia-Enceladus) have significantly less $\alpha$-elements than the red sequence stars (formed in the main progenitor). The run of [$\alpha$/Fe] vs. [Fe] in different stellar systems is commonly used to trace the star formation timescale of a galaxy, since $\alpha$ elements are produced in Supernova Type II on shorter timescales than iron, which is mostly produced on a longer timescale by Supernovae Type Ia. The fact that we find similar star formation timescales for the two components indicates that other mechanims should be at play. In our simulations, the progenitor and satellite have different star formation histories leading up to the merger \cite{fernandez_alvar2018}, with the progenitor star formation rate increasing whilst the satellite is small enough to have a basically constant star formation rate, regulated by feedback. 

\begin{figure}[h]
\begin{center}
\includegraphics[width=0.8\textwidth]{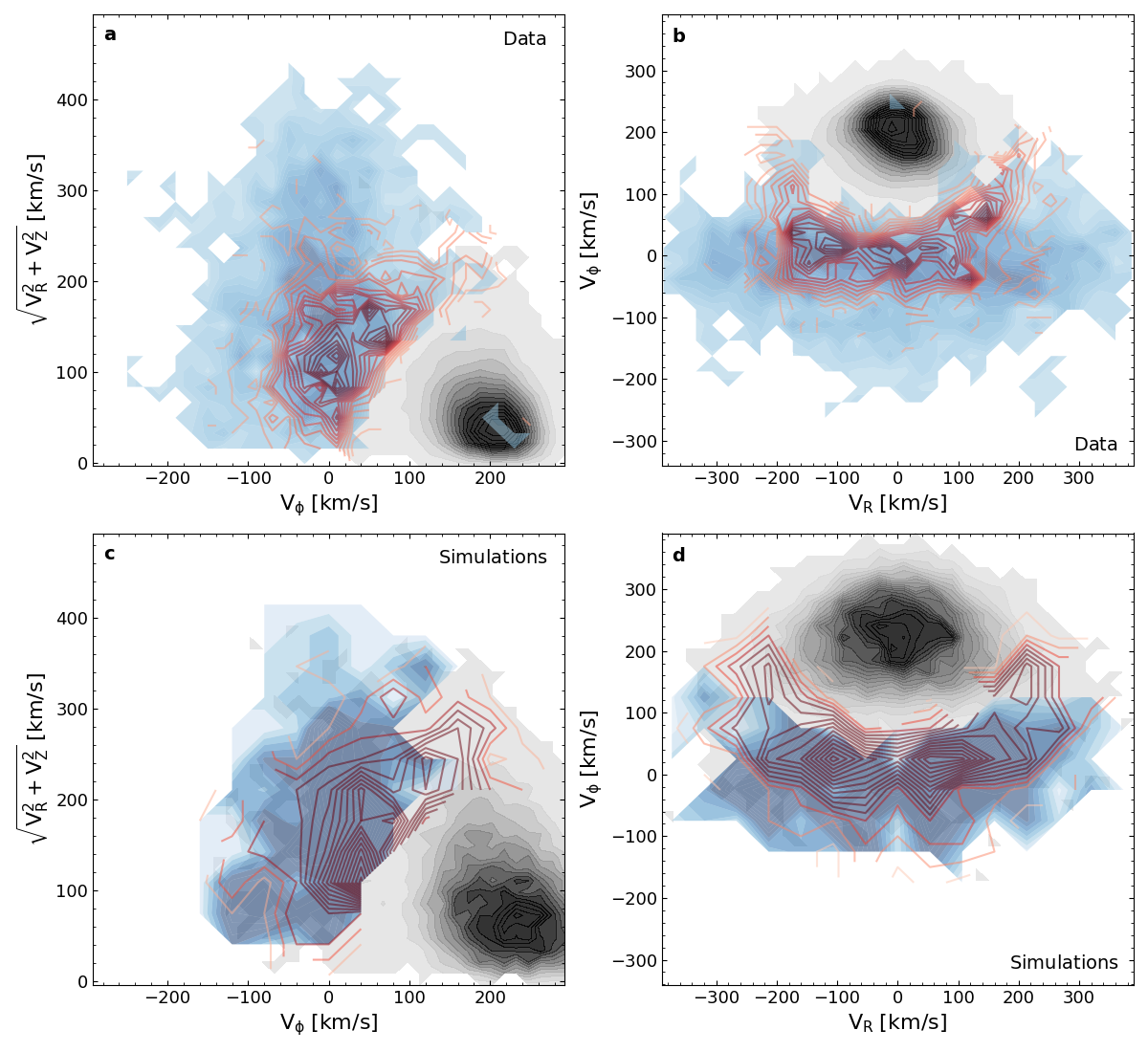} 
\end{center}
\caption{{\bf Comparison of the stellar kinematics between observations of Milky Way stars and the simulated Milky Way analogue.} {\bf a} and {\bf c} show the Toomre Diagram for Milky Way stars (obtained by combining Gaia DR2 astrometry and  LAMOST radial velocities\cite{sanders2018}), and for the simulated Milky Way analogue galaxy, respectively. {\bf b} and {\bf d} display similar information but in an azimuthal velocity (V$_{\rm \phi}$) versus radial velocity (V$_{\rm R}$) diagram. In the four panels, stars belonging to the observed halo blue sequence or the simulated accreted satellite are represented by blue shades, the observed halo red sequence or the simulated progenitor as red contours, and observed or simulated thick disc in grey scale. Note that some stars in our halo sample turn out to have disc properties when the full kinematic information is taken into account. This can help explain the presence of a small fraction of young stars in the halo sequences (panel a of Figure~\ref{fig:plot2}) as contamination from thin disc stars in the sample. The same is true for the inferred thick disc young population.
}

\label{fig:plot4}
\end{figure}

A further interesting result is the relation between the red sequence, or 'in-situ' halo, and the thick disc. The CMDs of these two components overlap, as seen in Figure~\ref{fig:plot1}, as do their ages, chemical abundances and kinematics as seen in Figures~\ref{fig:plot2} and~\ref{fig:plot4} as well as in previous studies\cite{babusiaux2018, haywood2018}.  The implication is that the in-situ halo and thick disc have the same origin, linked to the main progenitor of the present day Milky Way: in this self consistent scenario, the old age tail of the thick disc population shown in Figure~\ref{fig:plot2} is populated by  stars formed in  the main progenitor, which were not heated to halo-like velocities. We may speculate that the metal weak thick disk\cite{beers2014} would be part of this population. Therefore, stars that would be currently associated to the thick disc  were forming in the main progenitor before, during and after the merger with Gaia-Enceladus, whilst a  fraction of main progenitor stars formed very early, and were heated to halo kinematics by the accretion event. The infalling gas associated to the merger likely contributed to maintain and even stoke star formation in the early disk, which stellar age distribution shows a second peak of high star formation rate  $\simeq$ 9.5 Gyr ago. Such a formation scenario of the thick disc has been foreshadowed by cosmological galaxy formation simulations\cite{brook2004}: the thick disc forms at high redshift during a period characterised by gas rich mergers. 

A final implication of our findings regards the transition \cite{fuhrmann2011, recioblanco2014} between thick and thin disc formation. The derived ages for thick disc stars indicate that this transition occurred around 8-6 Gyr ago, in agreement with previous studies \cite{haywood2016}, and that this transition was not associated with the earlier Gaia-Enceladus merger. 

In conclusion, the reported new stellar age distributions, aided by state-of-the-art cosmological simulations of disc galaxy formation, delineate a clear picture of the formation of our Galaxy. In this picture, a primitive Milky Way had been forming stars during some 3 Gyr when a smaller galaxy, which had been forming stars on a similar timescale but was less chemically enriched owing to its lower mass, was accreted to it. This merger heated part of the existing stars in the main progenitor to a stellar halo-like configuration.  A ready supply of infalling gas during the merger ensures the maintenance of a disc-like configuration, with the thick disc continuing to form stars at a substantial rate. 
%Our measured age distributions indicate that the thick disc reached its peak star formation rate around 9 Gyr ago, that is 4.5 Gyr after the first stars formed in the Milky Way. 
Subsequently, around 8-6 Gyr ago,  the gas  settled into a thin disc that has continued to form stars until the present day.

\vspace{0.25cm}

\noindent
{\bf Correspondence and requests for materials} should be addressed to CG ({\it carme.gallart@iac.es}).

\vspace{0.25cm}

\vspace{0.25cm}

\noindent
{\bf Acknowledgements} CG, TRL and MM  acknowledge support by the Spanish Ministry of Economy and Competitiveness (MINECO) under the grants AYA2014-56795-P and AYA2017-89076-P as well as AYA2016-77237-C3-1-P (TRL). SC acknowledges support from Premiale INAF "MITIC" and has been supported by INFN (Iniziativa specifica TAsP). We used data from the European Space Agency mission Gaia (\url{http://www.cosmos.esa.int/gaia}), processed by the Gaia Data Processing and Analysis Consortium (DPAC; see \url{http://www.cosmos.esa.int/web/gaia/dpac/consortium}). Funding
for DPAC has been provided by national institutions, in particular the institutions participating in the Gaia Multilateral Agreement. We also used data from the LAMOST and GALAH surveys. Guoshoujing Telescope (the Large Sky Area Multi-Object Fiber Spectroscopic Telescope LAMOST) is a National Major Scientific Project built by the Chinese Academy of Sciences. Funding for the project has been provided by the National Development and Reform Commission. LAMOST is operated and managed by the National Astronomical Observatories, Chinese Academy of Sciences. The GALAH survey is based on observations made at the Australian Astronomical Observatory, under programmes A/2013B/13, A/2014A/25, A/2015A/19, A/2017A/18. We acknowledge the traditional owners of the land on which the AAT stands, the Gamilaraay people, and pay our respects to elders past and present.

\vspace{0.25cm}

\noindent
{\bf Author contributions} All authors have critically contributed to different aspects of the data analysis and model calculation, and to the interpretation of the results. The writing of the manuscript to which all authors contributed, was led by CG and CBB. Regarding specific aspects, CG selected the Gaia samples and performed the CMD fitting. EJB helped with the data selection, wrote the CMD fitting software ({\it TheStorm}) based on earlier work with MM and CG, and performed the calculation of the 3D interstellar reddening. CBB contributed the galaxy formation models which were key in the interpretation of the results. TRL participated in the CMD fitting, contributed key software for various steps such as the error simulation in the synthetic CMDs, and created the figures. SC contributed the software to calculate the synthetic CMD, including all the necessary libraries of stellar models and bolometric correction tables for the Gaia photometric passbands. VH selected the spectroscopic samples, which were analyzed together with TRL. 

\vspace{0.25cm}

\noindent
{\bf Competing interests} The authors declare no competing financial interests

\vspace{0.25cm}

\noindent
{\bf Correspondence and requests for materials} should be addressed to CG ({\it carme.gallart@iac.es}).

\vspace{0.25cm}

\vspace{0.25cm}

\vspace{0.25cm}

\vspace{0.25cm}

\newpage

\section*{Methods}

\section{Data selection}

A detailed and reliable determination of the age and metallicity distributions of a stellar system can be obtained from a CMD that is essentially complete down to the oldest main sequence turn-offs \cite{gallart2005}. For a relatively metal-rich stellar population such as that of the Milky Way, the old turnoff occurs  at an absolute magnitude M$_{\rm G}$=4 in the Gaia G band, but it is necessary to reach about one magnitude deeper to accurately sample this critical point in the CMD. At a distance of 2 Kpc from the Sun, M$_{\rm G}$=5 corresponds to an apparent magnitude G=16.5, down to which  Gaia DR2  is essentially complete.

From Gaia DR2\cite{brown2018} we selected all the stars inside a sphere of radius $\simeq$ 2 Kpc centered in the Sun down to M$_{\rm G,R}$=7 (where M$_{G,R}$  stands for M$_{\rm G}$ before applying any correction for interstellar extinction.  M$_{\rm G,R}$  was calculated directly from the apparent magnitude using the Gaia parallaxes, as M$_{\rm G,R}$=G+5+5${\rm \times}$log$_{10}$($\rm \omega$/1000.0), with ${\rm \omega}$ being the parallax in milliarcseconds (mas). A parallax zero-point offset of 0.054 \cite{Schonrich2019} mas (in the sense that the corrected parallaxes are larger), has been applied. This value has been preferred to the Gaia DR2 recommended value of 0.03 mas \cite{Lindegren2018} since several subsequent works have found a larger value \cite{Graczyk2019, Hall2019, Khan2019, LeungBovy2019, Muraveva2018, Zinn2018}.   The direct calculation of distance from the parallax value is valid because the great majority (over 95\%) of stars in our sample have small relative parallax error (\verb|parallax_over_error| $>$ 5 \cite{luri2018}).  The stars excluded by this cut are more scattered in the CMD than the thick disk and halo sequences. Even if they would correspond to actual stars in these populations, their numbers are so low that the CMD fit or the derived age distribution would not be affected. Finally, following \cite{babusiaux2018, Lindegren2018}, it was also required that \verb|astrometric_chi2_al|/(\verb|astrometric_n_good_obs_al| - 5) $<$ 1.44$\rm \times$max(1, exp(-0.4$\rm \times$ (\verb|phot_g_mean_mag|-19.5))) (only 2\% of stars were removed by this cut) and that the stars had (G$_{\rm BP}$-G$_{\rm RP}$) measured colour. 

From this parent population, we selected as thick disc sample all the stars in the North and South caps of the sphere, with distances from the Galactic plane, $|$Z$|$ $>$1.1 Kpc, and tangential velocities, calculated as V$_{\rm T}$=(4.74/$\rm \omega$)$\rm \times \sqrt{\verb|pmra|^2+\verb|pmdec|^2}$, lower than 200 km/s. For the halo sample, we selected all the relatively high Galactic latitude stars (b$>$30) with V$_{\rm T}>$200 km/s. The latitude cut was performed to avoid substantial contamination from the thin disc at low galactic latitude. Note that, since distances are actually obtained directly from parallax inversion, as d(pc)= 1000./\verb|parallax|, a small bias is introduced, such that true distances would tend to be smaller \cite{BailerJones2018} by $\simeq$ 100 pc on average. This bias affects very mildly the exact spatial shape and size of the actual volume covered, but does not change the fact that we are considering a thick disk sample (defined morphologically) and a halo sample (defined kinematically), and thus does not affect the conclusions of the paper. 

Corrections for interstellar extinction were calculated on a star by star basis by interpolating the colour excess from  a 3-D map \cite{lallement2018}. This map is strictly valid only up to Heliocentric distances of 2.6 Kpc and $|$Z$|$ $<$0.6 Kpc, and thus, our thick disk sample and part of the halo sample are outside its validity range. However, most of the dust is confined in a layer thinner than 0.6 Kpc and consequently, the extinction to the thick disk stars integrated up to this height is a very good approximation of the total extinction. We used published extinction coefficients \cite{casagrandevandenberg2018}, together with the effective temperature of each star (we neglected the dependence on metallicity). The temperature was estimated using a correlation between  G$_{\rm BP}$-G$_{\rm RP}$ colour and temperature T$_{\rm eff}$ determined from a large sample of Gaia DR2 stars with measured T$_{\rm eff}$. Relatively small extinction values are inferred for the bulk of stars in both samples: 96\% [93\%] of the stars have A$_{\rm G} < $ 0.2 and 79\% [70\%] have A$_{\rm G} < $ 0.1 for the thick disk [halo] sample.

After keeping only stars brighter than M$_{\rm G}$=5, the thick disc sample contains  518,968  stars and the halo sample contains 61,427 stars. The CMD of these stellar samples are displayed in the top panels of Figure~\ref{fig:plot1}.

\section{Modelling of the CMD}

In order to derive age and metallicity distributions of each population, the observed CMDs have been modelled through the comparison with synthetic CMDs following well stablished techniques \cite{gallart1999, hernandez1999, holtzman1999, dolphin2002, apariciogallart2004, apariciohidalgo2009, monelli2010, cignonitosi2010, ruizlara2018} that have been used for over 20 years to determine detailed star formation histories for Local Group galaxies \cite{gallart2005, tolstoy2009}. These galaxies are sufficiently close to observe individual stars, yet far enough away that all their stars can be considered to be at the same distance, which can be obtained accurately using various well calibrated distance indicators \cite{benedict2007, beaton2016}. 

However, applying this technique to populations of Milky Way stars had only been possible within an extremely small volume around the Sun \cite{hernandez2000, bertellinasi2001, cignoni2006, bernard2018proc} owing to the difficulty of obtaining accurate distances for large stellar samples. This is one of the many aspects in which the Gaia mission is providing a breakthrough: in its second data release (DR2)\cite{brown2018}, Gaia provided distances for 1.3 billion stars in the Milky Way. This distance information has allowed us, for the first time, to apply the CMD-fitting technique to the different morphological components of our own Galaxy and therefore to robustly derive their age distributions. This provides an unprecedented window to look at the early sequence of events that shaped the formation and structure of the Milky Way.

In the process of fitting and observed CMD, the combination of simple stellar populations (that is, synthetic populations with small ranges of ages, $\rm \Delta$T $\leq$ 1 Gyr, and metallicities, $\rm \Delta$ [Fe/H]$\sim$0.1 dex) which provide the best fit is obtained. The minimization has been performed with the code {\it TheStorm} \cite{bernard2015letter, bernard2018MNRAS}. This code uses a {\it mother} synthetic CMD from which the simple stellar populations are extracted. In order to check for possible systematic effects derived from the chemical properties of the stellar evolution models, these synthetic CMDs have been calculated using the BaSTI stellar evolution library in both its solar-scaled \cite{pietrinferni2004} and $\alpha$-enhanced heavy element distribution versions \cite{pietrinferni2006}. For the comparison itself, we make use of most of the area within the CMD populated by stars, excluding the Horizontal Branch (HB) which is still difficult to reproduce quantitatively in stellar evolution models. Although this portion of the CMD is not included in the fit, we tried to obtain a fine match of this evolutionary stage in the observed CMDs of both the halo -where it is quite prominent- and the thick disc -where the blue extension of the HB is basically missing. Since, for any given metallicity, the HB morphology is mainly driven by the mass loss efficiency during the red-giant branch stage, we used stellar model sets accounting for different efficiency of this process appropriate for each component: we used $\rm \eta$=0.2 for simulating the thick disc CMD and $\rm \eta$=0.4 for the halo CMD, where $\rm \eta$ is the free parameter used in combination with the Reimers' mass loss law in stellar modelling \cite{pietrinferni2006}.  Although different $\rm \eta$ values have been chosen for the different components under analysis, we should state that the results presented in this paper are not contingent upon this particular choice. Common to all {\it mother} synthetic CMDs (with 10$^{\rm 8}$ stars) are a constant star formation rate with flat age  and metallicity distributions  (14 Gyr to 30 Myr, and -2.6 $<$ [Fe/H] $<$ 0.1)  and the Kroupa initial mass function \cite{kroupa1993}. A binary fraction of 0.7 with  mass ratio q uniformly distributed between 0.1 and 1 has been adopted. Finally, stellar evolutionary model predictions have been transferred to the Gaia photometric system by using appropriate bolometric correction tables using the prescriptions for the DR2 catalogue \cite{evans2018}.  

The results displayed in Figures~\ref{fig:plot1}, \ref{fig:plot2}, and~\ref{fig:plot3} are obtained with the $\alpha$-enhanced models ($\alpha$/[Fe]=0.4 \cite{pietrinferni2006}, which are the most adequate for thick disc and halo stellar populations. According to the analysis\cite{haywood2018} of the metallicity and $\alpha$ abundances\cite{nissen_schuster2010} of a fraction of stars in the blue and red sequences of the halo CMD, red sequence stars have $\alpha$/[Fe] $\simeq$ 0.3 (with a mild dependence on metallicity), a value also typical for thick disk stars\cite{hayden2015}. Stars in the blue sequence have $\alpha$/[Fe] that range between $\alpha$/[Fe]$\simeq$0.4 at the lower metallicity extreme and   $\alpha$/[Fe]$\simeq$ 0.15 at the higher metallicity extreme ([Fe/H]$\simeq-0.8$ dex). However, we note that our conclusions, and in particular the age distributions, which are the most robust output of our CMD fitting process, are not affected by the choice of a scaled-solar heavy element distribution; we expect, therefore, that they would be even less affected by the small differences in the $\alpha$/[Fe] that would be required to match exactly the chemical characteristics of the different populations. A figure equivalent to \ref{fig:plot2} displaying the results obtained adopting a solar scaled heavy element distribution is provided as supplementary material to demonstrate this point. 

Prior to the comparison between the observed Gaia CMDs and the {\it mother} synthetic CMDs, it is necessary to perform a simulation of the observational errors in the synthetic CMD, in such a way that it can be consistently compared with the observations. We have considered two error sources, namely: the error in the absolute magnitudes introduced by the formal error in the parallax, and the error in the Gaia photometry. To model the first error source, which dominates the error in M$_{\rm G}$,  we have statistically reproduced the  distributions of \verb|parallax| and \verb|parallax_over_error| of the observed sample in the synthetic CMD (the \verb|parallax_over_error| has been updated from the values provided in the Gaia DR2 archive taking into account the parallax offset).  Then, the theoretical M$_{\rm G}$ values are transformed to account for observational errors by adding  a correction given by a Gaussian distribution centred at 0 and with a standard deviation given by $\rm \Delta$M$_{\rm G}$=2.17/\verb|parallax_over_error|. 
%In this way, the dependency of \verb|parallax_error| with colour and magnitude is implicitly taken into account in the simulation of the errors.
The maximum parallax errors in the samples used in this paper are of the order of 0.1 mas.  Similarly, the errors in the Gaia photometry have been implemented by adding a dispersion of 0.004 magnitudes to the (G$_{\rm BP}$-G$_{RP}$)  of all synthetic stars at all magnitudes \cite{evans2018}. We have not simulated uncertainties related to the extinction correction. These are expected to slightly smooth the CMD, and therefore the details in the star formation history. These uncertainties are of the order of 0.02-0.03 mag in E(B-V) over most of the sky, and so one order or magnitude smaller than the dispersion of the CMD due to the parallax uncertainty (which vary between 0.0 and ~0.3, with a typical value of 0.15).

The comparison between the observed and synthetic CMDs is performed using the number of stars in small -magnitude boxes. No a-priori constraint on the age-metallicity relation is adopted: {\it TheStorm} code solves simultaneously for age and metallicity within the range covered by the simple stellar populations. For more details on the fitting procedure and error calculation, the reader is referred to published references \cite{monelli2010, hidalgo2011, bernard2018MNRAS}. 

%At this point, it is interesting to note the fact that the global average systematic error of -0.03 mas in the parallax \cite{Lindegren2018} has not been taken into account in our absolute magnitude calculation. As discussed in the former references, the minimization is repeated numerous times for each sample after shifting the observed CMD in steps of colour and magnitude with respect to the synthetic CMDs, in order to account for uncertainties in photometric zero-points, distance, and mean reddening. Interestingly, the shift systematically found by the algorithm, in order to achieve the best fit, was of about 0.1 mag, that is, very approximately the expected shift since we did not take into account the systematic. This is quite remarkable, and supports that both the average zero point value of the parallax, and the zero point of the stellar evolution models are correct.

The results of this CMD fitting process are the star formation rate as a function of time and the age-metallicity distribution. From these, it is possible to compute a best-fit CMD  (displayed in the lower panels of Figure~\ref{fig:plot1}) which is then used to derive the age and metallicity distributions displayed in Figures~\ref{fig:plot2} and~\ref{fig:plot3} (panel a). In the case of the halo CMD, the fitting process has been performed in the CMD including both the blue and the red sequences. The age and metallicity distributions presented in Figure~\ref{fig:plot2} separately for each sequence have been derived by drawing a polygon enclosing each sequence in the observed CMD (avoiding the red clump where both sequences naturally overlap) and using this polygon to assign synthetic stars in the best fit CMD to either the blue or the red sequence. The boundary between the blue and red sequences corresponds to the fiducial line shown in Figure~\ref{fig:plot1}. We have tested that this particular delimitation has a negligible impact on the results presented here. 

%Together with the vertices of both blue and red sequence polygons, the coordinates of this line are included in the additional material for reproducibility purposes. 

\section{Cross-match with spectroscopic surveys} 

We used a homogeneous catalogue\cite{sanders2018} of distances, 3-D velocities, and ages for the cross-match between major Milky-Way spectroscopic surveys and Gaia DR2. We have used this catalogue,  together with the original Gaia DR2 information and the original [Fe/H] information  from LAMOST third data release A, F, G, K catalogue \url{{http://dr3.lamost.org/}} (\verb|feb| value) \cite{zhao2012} and GALAH first data release (\verb|fe_h| value)   \cite{buder2018}, to select stars matching exactly the same spatial and kinematical selection as our CMDs. The cross-identifications were performed requiring exact match of the corresponding identification numbers in the different catalogues (Gaia \verb|source_id|, LAMOST \verb|obsid| and GALAH \verb|gaia_dr2_id|). We selected these two surveys because i) LAMOST includes a significant number of stars on the main-sequence turnoff region of the CMD overlapping with our halo sample (2,588 stars after quality cuts, see below), and with our thick disk sample (31,810 stars). SEGUE also has a significant overlap in the main sequence region, but the survey target selection function somehow selected mostly stars on the blue main sequence. The other two high resolution surveys in the catalog (GES and APOGEE) contained too few stars in the main sequence turnoff region to be used here. While GALAH offers only a small sample of stars overlapping with our halo sample (282 stars after quality cut), it include main sequence stars covering both blue and red sequences (133 and 125 stars, respectively), and it provides a larger overlap with the thick disc sample (1460 stars), made essentially of giant stars. The collected spectroscopic samples were used in Figure 2 to build a metallicity distribution for the blue and red sequence halo, as well as the thick disc, and in Figure 4 to explore the 3D kinematics of the samples in cylindrical coordinates.

Some quality flags were used to clean the spectroscopic samples: both were required to have a quality flag=0 in the compilation catalogue \cite{sanders2018}, as advised by the authors. Additionally, LAMOST stars were considered only when \verb|snrg| $\geq$ 30 and GALAH stars were considered only when \verb|flag_cannon| = 0.

\section{Galaxy simulation}

Cosmological simulations of the Milky Way's halo have generally been done by simulating halos of similar mass to the Milky Way ($\sim 10^{12}M_\odot$), often imposing a relatively quiet merger history for the Milky Way since redshift $\sim$ 1. A range of increasingly detailed explorations  of the origins of high kinematic stars have been made using such simulations \cite{Brook04, Okamoto05, Font11, Pillepich14, Rodriguez-Gomez16, Tissera18,  Obreja18, Monachesi19, Fattahi19}. 

For this study, we had a suite of 18 Milky Way mass galaxies from the NIHAO \cite{wang15} and MaGICC \cite{brook2012magicc} programs, and selected the one which had a merger history and properties most like the Milky Way. The simulation used for  Figures~\ref{fig:plot3} and~\ref{fig:plot4} comes from a fully cosmological, hydrodynamical simulation of a Milky Way mass analogue galaxy from the MaGICC program. Importantly, these simulated galaxies have been shown to match a wide variety of galaxy scaling relations over a wide mass range, as well as reproducing a range of other galaxy properties such as enrichment of the Circum-Galactic Medium \cite{stinson2012}. The particular realisation, g15784, has total mass of 1.4$\rm \times$10$^{\rm 12}$ solar masses and has been shown to be in good agreement with the Milky Way in terms of its detailed chemical and kinematic properties of the thick and thin discs \cite{miranda2016}. In the present study we show that the stars in the region around the Sun match the most important properties of the stars observed in our own Milky Way, and use this to interpret the red and blue sequences of the observed halo CMD as belonging to the main Milky Way progenitor and Gaia-Enceladus respectively, with the merger of these two proto-galaxies resulting in the in-situ and vast majority of the inner stellar halo. 

In the simulation, stars form when gas has become sufficiently cool (T $<$ 10,000 K) and sufficiently dense (n$_{\rm th}$ $>$ 9.3 cm$^{\rm -3}$). When gas particles meet these criteria, stars form according to: dM$_{\rm *}$/dt = c$_{\rm *}$M$_{\rm gas}$/t$_{\rm dyn}$, where M$_{\rm *}$ is the mass of stars formed in time dt, M$_{\rm gas}$ is the mass of a gas particle, t$_{\rm dyn}$ is the dynamical time of gas particles, and c$_{\rm *}$ is the star formation efficiency -- i.e., the fraction of gas that will be converted into stars. Supernova feedback follows the blastwave model \cite{stinson2006} with thermal energy (10$^{51}$ erg) deposited to the surrounding ISM from each supernova. The simulations also include radiation feedback from massive stars \cite{stinson2013}. Cooling is disabled in the blast region ($\sim$ 100 pc) for $\sim$ 10 Myr. The simulation includes heating from a uniform UV ionising background radiation field \cite{haardtmadau1996}. Cooling takes into account both primordial gas and metals \cite{shen2010}. The metal cooling grid is derived using CLOUDY \cite{ferland1998}. 

Along with the fact that the same physical prescriptions used in this simulation resulted in matching scaling relations at a range of masses, meaning that low mass accreted satellites contain the correct amount of stars and gas for their halo mass, the correspondence with the observations of the thick disc gives us confidence in using this Milky Way analogue as an interpretive tool, as we do in this paper.

\end{document}